# Unraveling the Rank-Size Rule with Self-Similar Hierarchies

Yanguang Chen

(Department of Geography, College of Urban and Environmental Sciences, Peking University, 100871, Beijing, China. Email: chenyg@pku.edu.cn)

**Abstract**: Many scientists are interested in but puzzled by the various inverse power laws with a negative exponent 1 such as the rank-size rule. The rank-size rule is a very simple scaling law followed by many observations of the ubiquitous empirical patterns in physical and social systems. Where there is a rank-size distribution, there will be a hierarchy with cascade structure. However, the equivalence relation between the rank-size rule and the hierarchical scaling law remains to be mathematically demonstrated and empirically testified. In this paper, theoretical derivation, mathematical experiments, and empirical analysis are employed to show that the rank-size rule is equivalent in theory to the hierarchical scaling law (the $N^n$ principle). Abstracting an ordered set of quantities in the form $\{1, 1/2, \ldots, 1/k, \ldots\}$ from the rank-size rule, I prove a *geometric subdivision theorem* of the harmonic sequence ($k$=1, 2, 3, …). By the theorem, the rank-size distribution can be transformed into a self-similar hierarchy, thus a power law can be decomposed as a pair of exponential laws, and further the rank-size power law can be reconstructed as a hierarchical scaling law. A number of ubiquitous empirical observations and rules, including Zipf's law, Pareto's distribution, fractals, allometric scaling, $1/f$ noise, can be unified into the hierarchical framework. The self-similar hierarchy can provide us with a new perspective of looking at the inverse power law of nature or even how nature works.

# 1 Introduction

An attractive but mysterious phenomenon is that many types of data studied in the physical and social sciences can be approximated with the well-known Zipf distribution (Zipf, 1949). The numerical relations between rank and size generally follow Zipf's law, and the scaling exponent ($q$) is close to $q=1$ in most cases (Krugman, 1996). Another similar phenomenon is the $1/f^\beta$ noise, the numerical relation between frequency and spectral density always follow the inverse power law, and the spectral exponent ($\beta$) is often close to $\beta=1$. The rank-size distribution and the $1/f$-like noises occur widely in nature and human systems, and they are associated with fractals (Bak, 1996; Mandelbrot, 1999). A number of scientists are interested and puzzled by the rank-size pattern and $1/f$ noise, which appear in many complex systems. Despite numerous studies, the theoretical essence of these ubiquitous empirical observations has not yet been brought to light so far and remains to be explored.

The rank-size rule is often regarded as the special case of generalized Zipf's law related to the Yule distribution and the Pareto distribution. The Zipf distribution is in fact one of a family of varied scaling relationships (Bettencourt *et al*, 2007). If the scaling exponent ($q$) of the general Zipf distribution equals 1, the rank-size relationship is known as the rank-size rule (Knox and Marston, 2006). The rule describes a certain remarkable statistical regularity and forms a source of considerable interest in many fields. A great number of physical and social phenomena including the distribution of city sizes in a nation, sizes of business firms, wealth among individuals, lengths of rivers, areas of islands and islets, street hierarchies, particle sizes, and frequencies of word usage, satisfy the rank-size distribution empirically (Axtell, 2001; Batty, 2008; Berry, 1961; Blank and Solomon, 2000; Jiang, 2009; Konopka and Martindale, 1995; Marquet, 2002; Rybski *et al*, 2009; Simon and Bonini, 1958; Turcotte, 1997). If we can reveal the underlying rationale of the rank-size scaling relation, the related discrete power law probability distributions as well as $1/f$ noise will become more understandable.

In fact, many scientists study power laws, but little knows that the simplest approach to researching a power law is a pair of exponential laws. The key is the self-similar hierarchy. Davis (1978) once made an interesting discovery about hierarchies of cities. If we group the city sizes around the world into different classes according to the geometric sequence with the common ratio equal to 2, the city numbers in these classes also approximately form a geometric sequence and the common ratio approaches 2. The size-number relation of urban hierarchies is known the $2^n$ rule ($n=0, 1, 2, \ldots$). The $2^n$ principle is a theoretical window to probe into the rank-size rule and the similar power law distributions. This paper will discuss the following questions. First, where there is a rank-size rule, there is a $2^n$ rule and *vice versa*. The $2^n$ rule is actually an equivalent of the rank-size rule. Second, the $2^n$ rule can be generalized to the $3^n$ rule, the $4^n$ rule, and so on. Third, the hierarchies following the $2^n$ rule possess fractal structure and can be used to investigate many observations of the ubiquitous empirical patterns such as fractals and $1/f$ noise.

Hierarchy is frequently observed within the natural living world as well as in social institutions, representing a form of organization of complex systems without characteristic scales (Puman, 2006). An interesting discovery of this study is that the self-similar hierarchy is a magic framework, the macro property of which is independent of the elements at the micro level. The remainder of this article will be structured as follows. First, the rank-size rule will be abstracted as a harmonic sequence, from which the $2^n$ rule, $3^n$ rule, $4^n$ rule, and generally, the $N^n$ rule ($N$ refers to natural number) will be mathematically derived, and thus the **geometric subdivision theorem** of harmonic sequence is proved. In this process, the scaling relation between rank and size are decomposed as two exponential laws, and then reconstructed as a hierarchical power law. Then, three typical mathematical experiments and an empirical case are employed to confirm the theoretical derivation and hypothesis. Finally, the hierarchy models are generalized to encompass a number of ubiquitous general empirical observations.

## 2 Mathematical derivation (model)

### 2.1 Rearrangement of rank-size distribution

One of the typical rank-size distribution phenomena is cities, the things familiar to us. In urban studies, if the population size of a city is multiplied by its rank, the product will probably equal the population of the highest ranked city. If so, we will say the cities follow the rank-size rule. Mathematically, the rule states that the population of a city ranked $k$ will be $1/k$th of the size of the largest city $P_1$ (refer to *Oxford Dictionary of Geography*). The rule can be expressed as

$$P_k = \frac{P_1}{k}. \tag{1}$$

For simplicity, let $P_1$=1 unit. Then the rank-size rule can be abstracted as a harmonic sequence such as $\{1/k\}$, where $k$=1, 2, 3, …, and equation (1) suggests a special $1/k$ distribution. Now, group the harmonic sequence into different classes in terms of geometric sequence. Thus, a hierarchy of cities will be constructed by taking a primate city ($N^0$), generating $N$ next order cities, then $N^2$, and $N^3$, and so on ($N$ is a positive integer). For example, for $N$=3, the hierarchy of numbers indicating city sizes is illustrated in Figure 1, which only shows the first four classes as a schematic diagram.

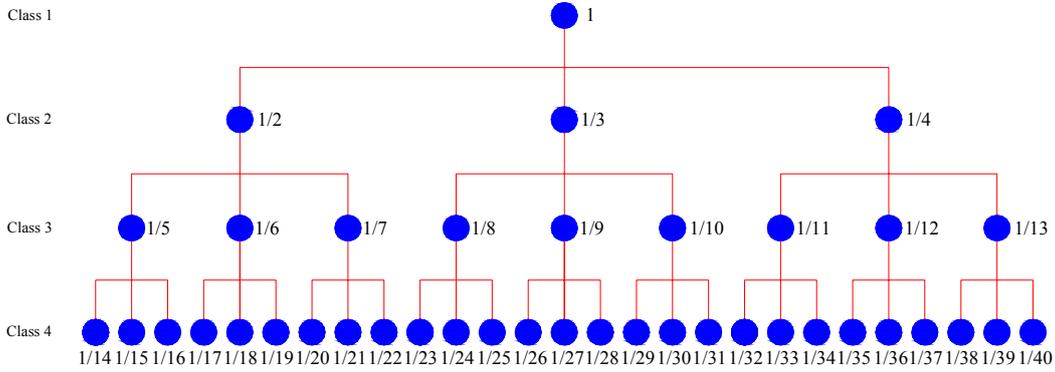

**Figure 1** The self-similar hierarchy of cities based on the harmonic sequence from the ideal rank-size distribution (The first 4 classes for $N$=3)

The city numbers in the hierarchy can be expressed as an exponential function. According to the mode of sequence subdivision, the number of order $m$, $f_m$, can be defined as

$$f_m = N^{m-1} = f_1 r_f^{m-1}, \qquad (2)$$

where $f_1$=1. Obviously, the interclass number ratio, $r_f$, equals the common ratio, $N$, that is

$$r_f = \frac{f_{m+1}}{f_m} = N. \qquad (3)$$

What interests us is the average value of each class. If we can prove that the sum of numbers in each level ($S_m$) approaches a constant asymptotically, that is $S_m=f_m P_m=constant$, then the average size of different classes ($P_m$) will decay exponentially, i.e., $P_m=constant/N^{m-1}$. This suggests that the rank-size rule is mathematically equivalent to the $N^n$ rule. In this instance, the rank-size distribution will be converted into a hierarchy with cascade structure and fractal dimension.

## 2.2 The geometric subdivision theorem of the harmonic sequence

The above question can be abstracted as a mathematical problem to be proved, and the proof is very simple. According to the rule of subdivision of harmonic sequence based on geometric proportion, the summation of numbers in each class can be expressed as

$$S_m = \frac{1}{\frac{1}{N-1}(N^{m-1}+N-2)} + \frac{1}{\frac{1}{N-1}(N^{m-1}+N-2)+1} + \cdots + \frac{1}{\frac{1}{N-1}(N^{m-1}+N-2)+N^{m-1}-1}$$
$$= \sum_{j=0}^{N^{m-1}-1} \frac{1}{\frac{1}{N-1}(N^{m-1}+N-2)+j}, \quad (4)$$

where $S_m$ is the number sum of order $m$, $m=1, 2, 3, \ldots M$, $N=2, 3, 4\ldots$, $j=0, 1, 2, \ldots, N^{m-1}-1$. Here $M$ is the largest number of class order. Mathematical experiments show that the sum of numbers in each level will approach $\ln N$ as $m=M\to\infty$ (see Subsection 3.1). If we can prove

$$\lim_{m\to\infty} S_m = \ln N, \quad (5)$$

then the above problem will be resolved, and the conclusion can be obtained that the rank-size rule can be theoretically led to the $N^n$ rule.

By equation (4), the number sum of each class can be rewritten in the form

$$S_m = \frac{\frac{N-1}{N^{m-1}+N-2}}{1+\frac{N-1}{N^{m-1}+N-2}\cdot 0} + \frac{\frac{N-1}{N^{m-1}+N-2}}{1+\frac{N-1}{N^{m-1}+N-2}\cdot 1} + \cdots + \frac{\frac{N-1}{N^{m-1}+N-2}}{1+\frac{(N-1)}{N^{m-1}+N+2}\cdot(N^{m-1}-1)}$$
$$= \sum_{j=0}^{N^{m-1}-1} \frac{\frac{N-1}{N^{m-1}+N-2}}{1+\frac{N-1}{N^{m-1}+N-2}\cdot j}. \quad (6)$$

Suppose the function $f(x)=1/(1+x)$ is defined within the interval $[0, N-1]$. Using the fractions

$$0 = \frac{N-1}{N^{m-1}+N-2}\times 0, \frac{N-1}{N^{m-1}+N-2}\times 1, \cdots, \frac{N-1}{N^{m-1}+N-2}\times(N^{m-1}+N-2) = N-1$$

to divide the closed interval $[0, N-1]$ into $N^{m-1}+N-2$ equal parts with a length of $(N-1)/(N^{m-1}+N-2)$. Further defining

$$x_j = \frac{N-1}{N^{m-1}+N-2}\times j, \quad (7)$$

we have

$$\Delta x_j = \frac{N-1}{N^{m-1}+N-2}. \quad (8)$$

Thus the sum of the $m$th class can be expressed as

$$\lim_{m\to\infty} S_m = \lim_{m\to\infty} \sum_{j=0}^{N^{m-1}-1} \frac{\frac{N-1}{N^{m-1}+N-2}}{1+\frac{N-1}{N^{m-1}+N-2}\cdot j} = \lim_{m\to\infty} \sum_{j=0}^{N^{m-1}-1} \frac{\Delta x_j}{1+x_j}, \quad (9)$$

Apparently, the final result of summation is

$$\lim_{m \to \infty} S_m = \lim_{m \to \infty} \sum_{j=0}^{N^{m-1}-1} f(x_j) \Delta x_j = \int_0^{N-1} \frac{1}{1+x} dx = [\ln(1+x)]_0^{N-1} = \ln N. \tag{10}$$

This indicates $S_m$ is asymptotically close to $\ln N$ as $m \to \infty$, and the proof of equation (5) is complete.

The mathematical proof gives the **_geometric subdivision theorem_** of the harmonic sequence: If we group a harmonic sequence $\{1/k\}$ into different classes, and the amount of numbers in each class form a geometric sequence such as $N^0$, $N^1$, $N^2$, ..., $N^m$, then the sum of numerical value in each class asymptotically approaches a constant $\ln N$.

An exponential law of average size can be derived from the geometric subdivision theorem. Defining the average size in the $m$th class as $P_m = S_m/f_m$ yields $S_m = f_m P_m$. According to equation (5), under the limit condition, we have

$$f_m P_m = \ln N. \tag{11}$$

This implies a special inverse power law $f_m = (\ln N) P_m^{-1}$. In the light of equation (11), the average size ratio of adjacent classes is

$$r_p = \lim_{m \to \infty} \frac{P_m}{P_{m+1}} = \frac{f_{m+1}}{f_m} = N, \tag{12}$$

in which the common ratio $r_p$ denotes the size ratio. The rank-size relationship suggests a fractal structure (Mandelbrot, 1983; Chen and Zhou, 2006), and the similarity dimension of the hierarchy is

$$D = \frac{\ln r_f}{\ln r_p} = \frac{\ln f_{m+1}/f_m}{\ln P_m/P_{m+1}} \to \frac{\ln N}{\ln N} = 1, \tag{13}$$

An exponential model can be deduced from equation (12) by recursion, and the result is

$$P_m = P_{m-1}/N = N^{1-m} = P_1 r_p^{1-m}. \tag{14}$$

So far, the inverse power law, equation (1), has been decomposed as a pair of exponential laws: one is the number law, equation (2), and the other is size law, equation (14). If the former is defined, we will have the latter, and *vice versa*. The exponential functions can be termed "the $N^n$ rule" of hierarchies. When $N=2$, the exponential laws is just the $2^n$ rule (Davis, 1978). From equations (2) and (14) follows that

$$f_m = \mu P_m^{-D}, \tag{15}$$

where $\mu=f_1P_1=1$ refers to the proportional coefficient, and $D=\ln r_f/\ln r_p=1$ to the fractal parameter defined by equation (13). This suggests that the rank-size scaling law of size distributions has been transformed into the size-number scaling law of hierarchical structure.

In theory, if $m=1$ and $D=1$ as given, then $\mu= f_1P_1=1\neq \ln N$. However, comparing equation (15) with equation (11) shows that $\mu=S_m=\ln N$. This implies that if the harmonic sequence is rearranged by a geometric sequence, the fractal structure of the hierarchical systems comes into being gradually. Fractal dimension is a parameter under limit condition, and the first class or the largest city always makes an exception. This is consistent with the definition of fractals (Mandelbrot, 1983). In practice, if we apply the scaling law to the real hierarchies of cities, the data points of size *vs* number may depart from the scaling range on the log-log plot when $m$ is very small (esp. $m=1$).

Now, the geometric subdivision theorem of the harmonic sequence can be re-expressed as follows. If a harmonic sequence $\{1/k\}$ is arranged into a hierarchy with cascade structure in terms of a geometric sequence, $N^0, N^1, N^2, …, N^m$, then the average values in different classes will decay in a negative exponential way. This theorem indicates one of the magic properties of the harmonic sequence.

## 3 Mathematical experiments and empirical evidence (result)

### 3.1 Mathematical experiments

A number of mathematical experiments can be performed to verify the geometric subdivision theorem of the harmonic sequence, and thus to support the equivalent relationship between the rank-size distribution and the self-similar hierarchical structure. The mathematical experiment is so simple that we can carry out it with MS Excel. Let's take city-size distribution as an example, and let $P_1=1$ unit to represent the size of the largest city. Under the ideal condition, the city sizes form an harmonic sequence such as $[1, 1/2, 1/3, …, 1/k, …]$ in terms of equation (1). If $N=2$ as given, then the hierarchy of cities is as follows: $[1]$; $[1/2, 1/3]$; $[1/4, 1/5, 1/6, 1/7]$;…; $[1/(2^{m-1}), 1/(2^{m-1}+1),…,1/(2^m-1)]$;…. The common ratio of city numbers in different classes is $r_f=f_{m+1}/f_m=2$, where $m=1, 2, …, M$, and the positive integer $M$ approaches infinity in theory. In these experiments, we can take $M=10$, which is enough to show the hierarchical regularity.

Now, let's see the sum of city sizes in each class. For the first class, the sum is 1; for the second class, the sum is 1/2+1/3≈0.8333; for the third class, the sum is 1/4+1/5+1/6+1/7≈0.7595, and so on. When the class order $m$ becomes larger and larger, the sum is closer and closer to ln(2)≈0.6931. The average size of each class $P_m$ approaches $\ln(2)/2^{m-1}$, so the average size ratio $r_p=P_m/P_{m+1}$ approaches $r_f=f_{m+1}/f_m=2$. Thus the fractal dimension goes near $D=1$. For the first 10 classes, the results are listed in Table 1 and displayed in Figure 2(a). The estimated value of the fractal dimension is about $D=0.9559$. If $M$ is large enough, or the first class is removed as an outlier in the least squares computation, the dimension value will be adequately close to 1.

**Table 1** The results of mathematical experiments by converting the harmonic sequence based on the rank-size rule into the geometric sequences (examples)

| Common ratio ($r_f$) | Class ($m$) | City number ($f_m$) | Total size ($f_m P_m$) | Average size ($P_m$) | Size ratio ($r_p$) |
|---|---|---|---|---|---|
| $N=2$ | 1 | 1 | 1 | 1 | |
| | 2 | 2 | 0.83333 | 0.41667 | 2.40000 |
| | 3 | 4 | 0.75952 | 0.18988 | 2.19436 |
| | 4 | 8 | 0.72537 | 0.09067 | 2.09416 |
| | 5 | 16 | 0.70902 | 0.04431 | 2.04614 |
| | 6 | 32 | 0.70102 | 0.02191 | 2.02281 |
| | 7 | 64 | 0.69707 | 0.01089 | 2.01134 |
| | 8 | 128 | 0.69510 | 0.00543 | 2.00565 |
| | 9 | 256 | 0.69412 | 0.00271 | 2.00282 |
| | 10 | 512 | 0.69364 | 0.00135 | 2.00141 |
| | … | … | … | … | … |
| | $M$ | $2^{M-1}$ | ln(2) | $\ln(2)/(2^{M-1})$ | 2 |
| $N=3$ | 1 | 1 | 1 | 1 | |
| | 2 | 3 | 1.08333 | 0.36111 | 2.76923 |
| | 3 | 9 | 1.09680 | 0.12187 | 2.96316 |
| | 4 | 27 | 1.09841 | 0.04068 | 2.99561 |
| | 5 | 81 | 1.09859 | 0.01356 | 2.99951 |
| | 6 | 243 | 1.09861 | 0.00452 | 2.99995 |
| | 7 | 729 | 1.09861 | 0.00151 | 2.99999 |
| | 8 | 2187 | 1.09861 | 0.00050 | 3.00000 |
| | 9 | 6561 | 1.09861 | 0.00017 | 3.00000 |
| | 10 | 19683 | 1.09861 | 0.00006 | 3.00000 |
| | … | … | … | … | … |
| | $M$ | $3^{M-1}$ | ln(3) | $\ln(3)/(3^{M-1})$ | 3 |
| | 1 | 1 | 1 | 1 | |
| | 2 | 4 | 1.28333 | 0.32083 | 3.11688 |
| | 3 | 16 | 1.36203 | 0.08513 | 3.76890 |
| | 4 | 64 | 1.38038 | 0.02157 | 3.94682 |
| | 5 | 256 | 1.38483 | 0.00541 | 3.98716 |
| $N=4$ | 6 | 1024 | 1.38593 | 0.00135 | 3.99682 |
| | 7 | 4096 | 1.38620 | 0.00034 | 3.99921 |
| | 8 | 16384 | 1.38627 | 0.00008 | 3.99980 |
| | 9 | 65536 | 1.38629 | 0.00002 | 3.99995 |
| | 10 | 262144 | 1.38629 | 0.00001 | 3.99999 |
| | … | … | … | … | … |
| | $M$ | $4^{M-1}$ | $\ln(4)$ | $\ln(4)/(4^{M-1})$ | 4 |

The second typical experiment is for $N=3$, and the hierarchy is [1]; [1/2, 1/3, 1/4]; [1/5, 1/6, …, 1/13];…. The organized mode and hierarchical structure are illustrated in Figure 1. For the first class, the sum is 1; for the second class, the sum is 1/2+1/3+1/4≈1.0833; for the third class, the sum is 1/5+1/6+…+1/13≈1.0968, and so on. The sum approaches $\ln(3)≈1.0986$ rapidly, and the average size runs to $P_m=\ln(3)/3^{m-1}$ swiftly. As a result, the average size ratio $r_p=P_m/P_{m+1}$ tends to $r_f = f_{m+1}/f_m=3$. For the first 10 classes, the sequences are listed in Table 1 and displayed in Figure 2(b). The estimated value of the fractal dimension is about $D=1.0052$, very close to the expected value 1.

The third experiment is for $N=4$, and the hierarchy is [1]; [1/2, 1/3, 1/4, 1/5]; [1/6, 1/7, …, 1/21];…. The process can be understood by the similar approach to the above ones, and the results are listed in Table 1 and displayed in Figure 2(c). The estimated value of the fractal dimension is about $D=1.0155$, near 1. The rest, i.e., for $N=5, 6, 7,…$, may be deduced by analogy. Theoretically, $N$ is an arbitrary positive integer. Of course, all these experiment results are approximate since only the first 10 classes are taken into account.

The hierarch of cities can be described with the size-number inverse power law. For example, for $N=3$, a least squares calculation based on the first 10 classes yields

$$\hat{f}_m = 1.0593 P_m^{-1.0052}.$$

The goodness of fit is $R^2=0.9999$ (Figure 2(b)). The power law can be decomposed as two exponential laws. The first model is *ad hoc* given as $f_m=3^{m-1}$, while the second one based on the first 10 classes is such as

$$\hat{P}_m = 3.1579e^{-1.0928m} = 1.0587 \times 2.9827^{1-m}.$$

The goodness of fit is $R^2=0.9999$, and the fractal dimension is estimated as $D \approx \ln(3)/\ln(2.9827) \approx$ 1.0053, very close to the value from the power function.

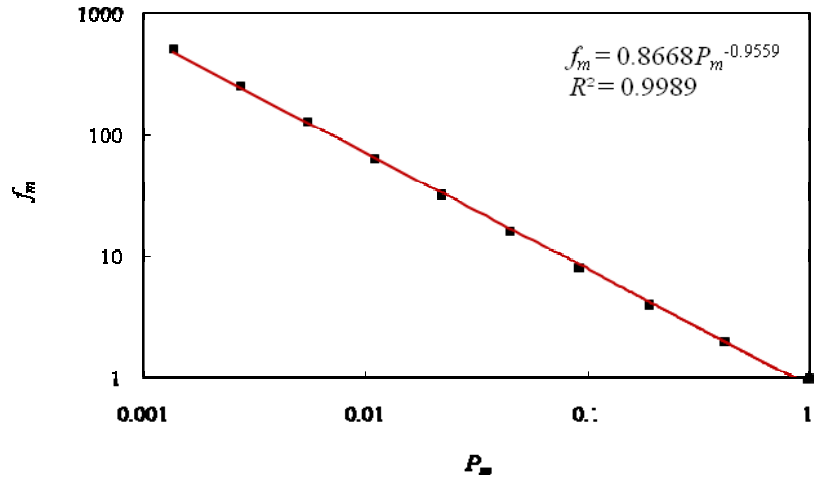

a. $N=2$

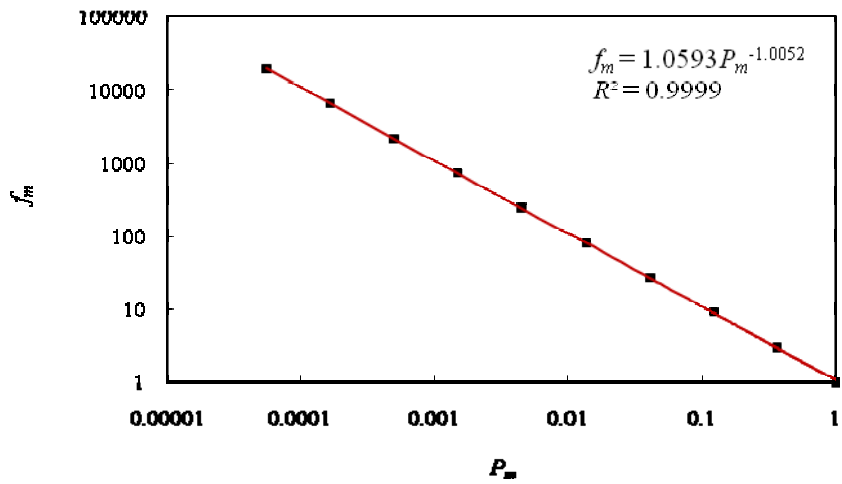

b. $N=3$

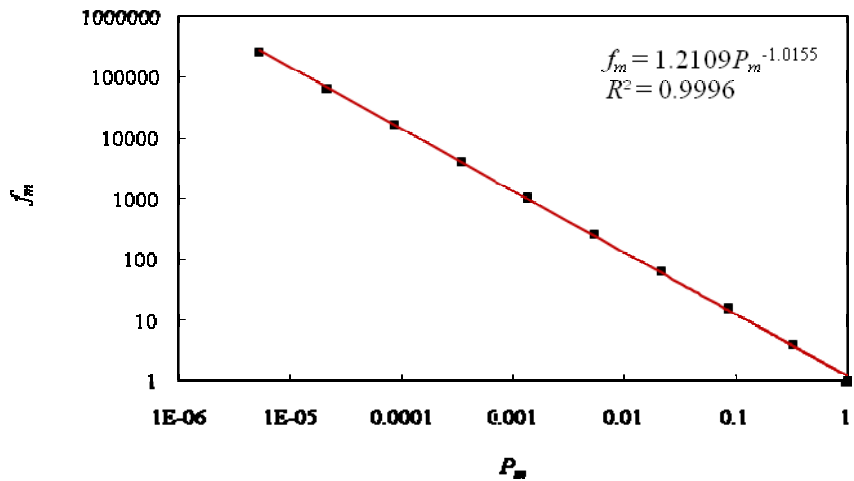

c. $N=4$

**Figure 2** The scaling relations between city numbers and average sizes of different classes in the ideal self-similar hierarchy of cities

## 3.2 Empirical evidence

Among various cases, cities in a given region are the typical phenomena which follow the rank-size scaling law (Batty, 2008; Berry, 1961; Carroll, 1982; Gabaix, 1999). If the region studied is large enough to cover the territorial scope influenced by the largest city, the city size distribution will comply with Zipf's law. If the scaling exponent is about $q=1$, we have a rank-size pattern. The cities of the United States of America (USA) will be employed to testify the models given above. A question is that the data of city population do not refer to units defined in strictly comparable terms. In urban geography, there three basic concepts used to define urban areas and populations, namely, city proper (CP) without suburb, urban agglomeration (UA) with suburb, and metropolitan area (MA) more suburbs (Davis, 1978). No matter what kind of urban area is considered, the city sizes satisfy the Zipf distribution in the case of very large scale. However, only the size distribution of the population defined within UA bears a scaling exponent near $q=1$, and on the whole, UA corresponds to concept of urbanized area.

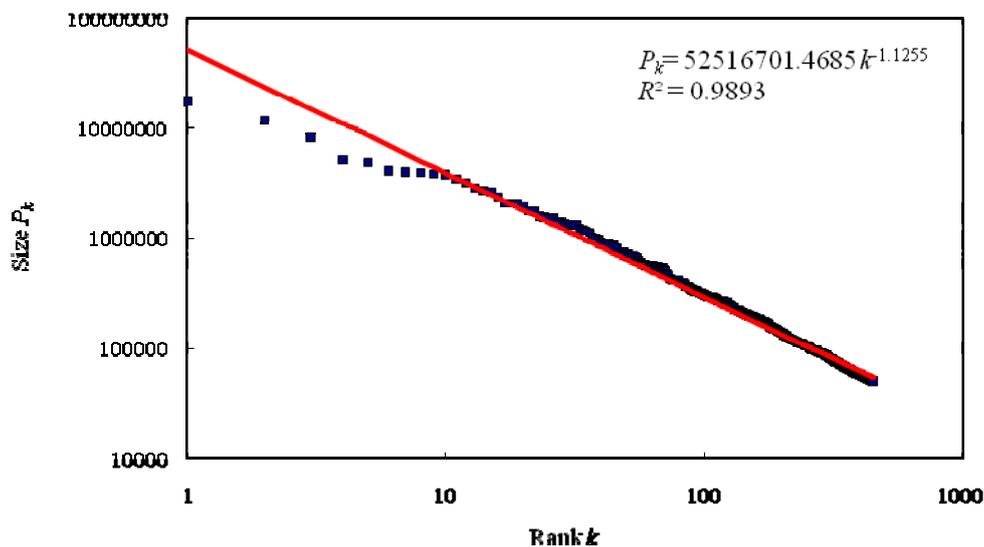

**Figure 3** The rank-size pattern of the US cities by the population within urbanized area (2000)

The data of this case study come from the US Census Bureau, but only the 452 cities with population in urbanized area are over 50,000 in 2000 and the data are available at internet

([www.demographia.com](www.demographia.com)). The Zipf model of these cities is such as

$$\hat{P}_k = 52516701.468 k^{-1.125},$$

The goodness of fit is about $R^2=0.989$, the scaling exponent is about $q=1.125$, and the corresponding fractal dimension can be estimated as $D=1/q \approx 0.9$. The shortcoming of the US case lies in two aspects: one is that the sample is not large enough (only 452 cities), the other is that the rank-size pattern is not so satisfying (Figure 3). Despite these flaws, the example is enough for us to illustrate the equivalence relation between the rank-size rule and the $N^n$ rule on self-similar hierarchical structure.

Corresponding to the mathematical experiments, three kinds of self-similar hierarchies will be constructed. Taking number ratio $r_f=2, 3, 4$, we can group the cities into different classes according to the $2^n$ rule, $3^n$ rule, and $4^n$ rule. The results, including city number, total population, and average population size in each class and size ratio between any two immediate classes, are listed in Table 2. In each hierarchy, two classes, i.e., top class and bottom class, are special and can be considered to be exceptional values. As can be seen in Tables 1 and 2, the first class always gets departure from the scaling range; the last class is always a *lame duck class* due to undergrowth or absence of data of small cities (Davis, 1978). For example, for $N=2$, the city number in the 9th class is expected to be $f_9=2^8=256$, but only 197 city data are available. In this instance, the first and last classes can be treated as outliers and should be removed from the least squares calculations. The data points of the median part form a scaling range on each log-log plot.

**Table 2** The results of empirical analysis by grouping the US city sizes (by urbanized area) in 2000 into different classes according as the $2^n$, $3^n$, and $4^n$ rules

| Common ratio ($r_f$) | Class ($m$) | City number ($f_m$) | Total population ($f_m P_m$) | Average size ($P_m$) | Size ratio ($r_p$) |
|---|---|---|---|---|---|
| | 1 | 1 | 17799861 | 17799861 | |
| | 2 | 2 | 20097391 | 10048696 | 1.771 |
| | 3 | 4 | 18246258 | 4561565 | 2.203 |
| | 4 | 8 | 26681941 | 3335243 | 1.368 |
| $N=2$ | 5 | 16 | 27052740 | 1690796 | 1.973 |
| | 6 | 32 | 26098069 | 815565 | 2.073 |
| | 7 | 64 | 22690390 | 354537 | 2.300 |
| | 8 | 128 | 19988240 | 156158 | 2.270 |
| | 9* | 197 | 13738825 | 69740 | 2.239 |
|     | 1   | 1   | 17799861 | 17799861 |       |
|     | 2   | 3   | 25246470 | 8415490  | 2.115 |
|     | 3   | 9   | 34392479 | 3821387  | 2.202 |
| N=3 | 4   | 27  | 42292382 | 1566385  | 2.440 |
|     | 5   | 81  | 37396956 | 461691   | 3.393 |
|     | 6   | 243 | 30338488 | 124850   | 3.698 |
|     | 7*  | 88  | 4927079  | 55990    | 2.230 |
|     | 1   | 1   | 17799861 | 17799861 |       |
|     | 2   | 4   | 30165506 | 7541377  | 2.360 |
| N=4 | 3   | 16  | 47236568 | 2952286  | 2.554 |
|     | 4   | 64  | 50896854 | 795263   | 3.712 |
|     | 5   | 256 | 39836606 | 155612   | 5.111 |
|     | 6*  | 111 | 6458320  | 58183    | 2.675 |

**Source**: The original data come from the US Census Bureau, available at: www.demographia.com. ***Note**: The last class of each hierarchy is actually a lame-duck class of Davis (1978).

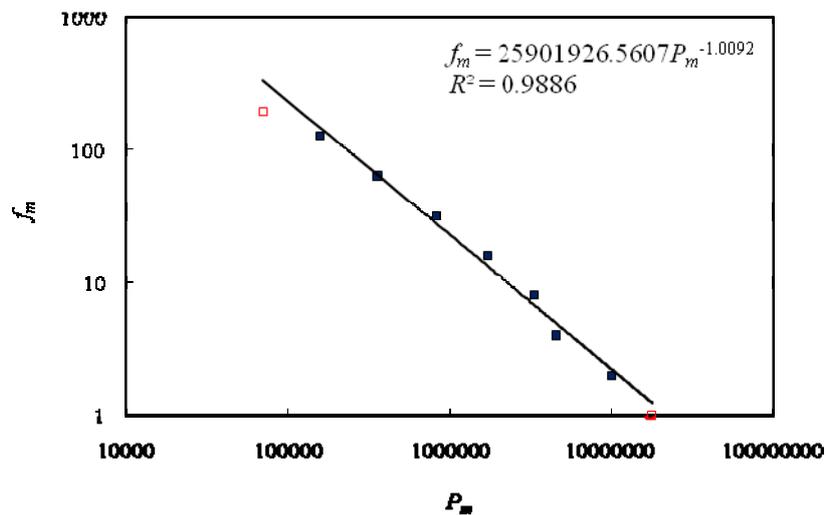

a. $N=2$

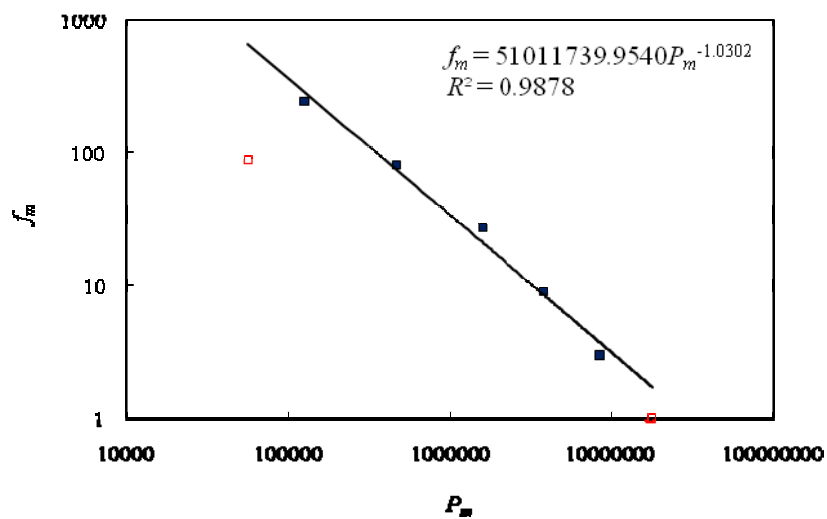

b. $N=3$

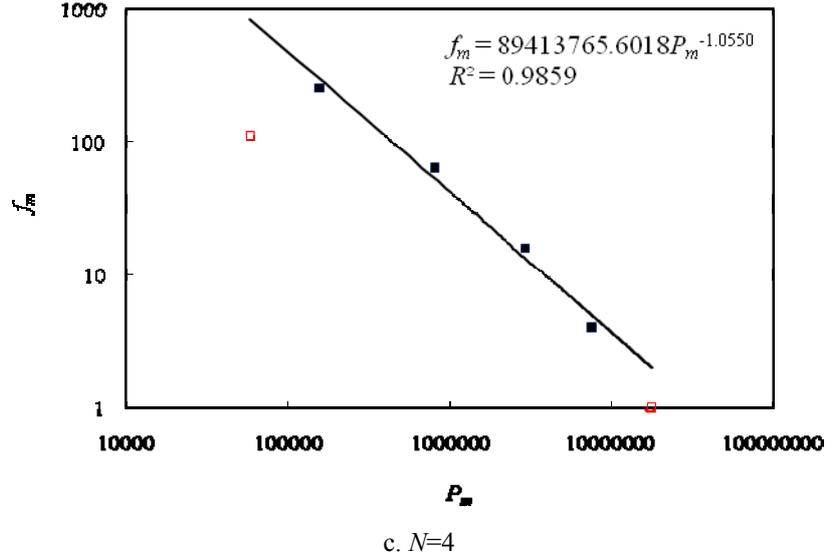

c. *N*=4

**Figure 4** The scaling relations between city numbers and average sizes in different classes of hierarchies of USA cities by the population in urbanized area in 2000 (examples)

By means of regression analysis, three power law models can be built (Figure 4). For *N*=2, the scaling relation is

$$\hat{f}_m = 25901926.561 P_m^{-1.009}.$$

The goodness of fit is about $R^2$=0.989, the fractal dimension is estimated as $D$≈1.009. The average size ratio within the scaling range is about $r_p$=2.031, which is very close to $r_f$=2. For *N*=3, the scaling relation is

$$\hat{f}_m = 51011739.954 P_m^{-1.030}.$$

The goodness of fit is about $R^2$=0.988, the fractal dimension is about $D$=1.030. The average size ratio within the scaling range is estimated as $r_p$=2.933, near $r_f$=3. For *N*=4, the scaling relation is

$$\hat{f}_m = 89413765.602 P_m^{-1.055}.$$

The goodness of fit is about $R^2$=0.986, the fractal dimension is estimated as $D$≈1.055. The average size ratio within the scaling range is about $r_p$=3.792, close to $r_f$=4.

Obviously, if we change the number ratio $r_f$, the average size ratio $r_p$ will change with it. However, the fractal dimension keep around $D$=1, which suggests the scaling exponent $q\to 1$. If the sample were large enough, we would take *N*=5, 6, 7, …. But for out empirical analysis, this example is sufficient to support the hierarchical scaling models. As expected, the hierarchies can be described with three pairs of exponential laws. For $N=2$, the number law is known as $f_m=(1/2)e^{\ln(2)m}$, the size law can be estimated as

$$\hat{P}_m = 41622813.522 e^{-0.686m}.$$

The goodness of fit is about $R^2=0.991$, and the fractal dimension can be indirectly estimated as $D\approx \ln(2)/0.686\approx 1.010$. For $N=3$, the number law is $f_m=(1/3)e^{\ln(3)m}$. If we remove the first class as an outlier, the size law can be estimated as

$$\hat{P}_m = 81154034.287 e^{-1.044m}.$$

The goodness of fit is about $R^2=0.993$, and the fractal dimension is about $D\approx \ln(3)/1.044\approx 1.052$. For $N=4$, the number law is $f_m=(1/4)e^{\ln(4)m}$. After removing the first class as an exceptional value, the size law can be estimated as

$$\hat{P}_m = 110245635.879 e^{-1.267m}.$$

The goodness of fit is about $R^2=0.992$, and the fractal dimension is about $D\approx \ln(4)/1.267\approx 1.094$.

These exponential models and the corresponding power models can be transformed into one another. From these exponential functions, the fractal parameters and the scaling exponents of the hierarchies can be evaluated approximated. All these parameter values of the self-similar hierarchical structure are close to the corresponding parameters values of the rank-size distributions. This suggests that the hierarchy with cascade structure can be described with exponential laws and power laws from different angle of views respectively. The $N^n$ rule can be applied to the cities in other countries such as Britain, France, and German (more empirical evidence is shown in an attached material).

## 4 Theoretical explanation and generalization (discussion)

The mathematical experiments and empirical analysis lend support to the inference that the rank-size rule is asymptotically equivalent to the $N^n$ rule as $m$ is large enough. But what is the significance of this study? The first is physical explanation, and the second the theoretical generalization. There are various explanations but no convincing explanation for the rank-size rule, especially in urban studies, despite the great frequency with which it has been observed. These years, some interesting explanations appeared in different fields (Bettencourt *et al*, 2007; Ferrer i

Cancho *et al*, 2005; Ferrer i Cancho and Solé, 2003). The rank-size distribution seems to be associated with maximum entropy models (Mora, 2010). In fact, the self-similar hierarchy is a simple link between the maximum entropy principle and the rank-size rule. First, equations (2) and (14) can be derived using the entropy-maximizing method (Chen and Liu, 2002; Curry, 1964). Second, combining equation (2) with equation (14) yields equation (15). Third, according to the ***geometric subdivision theorem*** of harmonic sequence, equation (15) is equivalent to equation (1). So, this suggests that the rank-size pattern comes from the process of entropy maximization, which is very important in urban modeling (Wilson, 1968; Wilson, 1971). The notion of entropy maximization of human systems is different from the concept of entropy increase in thermodynamics. A preliminary study suggests that the maximum entropy implies a compromise between the equity for individuals and the efficiency of the whole where social and economic systems are concerned.

The hierarchical structure can be employed to unify many different scaling rules and phenomena, and thus form an entire logic framework for us to understand nature. The rank-size rule suggests a magic sequence, which thus suggests a magic framework – a hierarchy with cascade structure. In a sense, the self-similar hierarchy can be compared to the hat of a magician, by which nature "produces" many "rules" and patterns such as the rank-size rule, the $2^n$ rule, the $3^n$ rule, Pareto distribution, Zipf's law, fractals, allometric growth, and 1/*f* noise (Figure 5). The relationships between simplicity and complexity of physical and social systems can been seen from these principles and patterns. Among these, fractals, 1/*f* noise, Zipf's law, and scale-free network are regarded as the ubiquitous general empirical observations (Bak, 1996; Barabási, 2002). Another magic property of this sequence is the latent Pareto distribution. Defining a size scale as $\xi=P_k=1/N^{m-1}$ ($k=N^{m-1}$; $N=2, 3, 4, \ldots$; $m=1, 2, 3, \ldots$), the number of the cities with size greater than or equal to $\xi$ is just $N(P\geq\xi)=N^{m-1}$. Apparently, we have an inverse power-law relation such as

$$N(P \geq \xi) \propto \xi^{-1}. \tag{16}$$

This suggests a special Pareto distribution with scaling exponent equal to 1. The scaling relation also demonstrates the fractal nature of the hierarchical structure and the rank-size distribution.

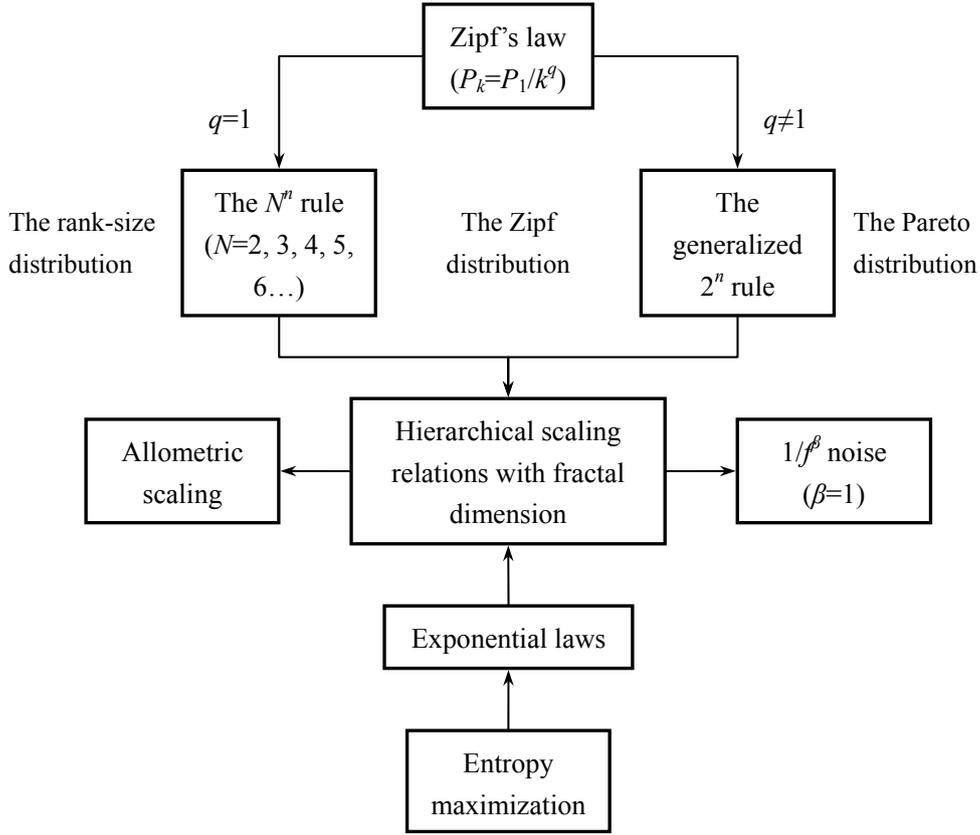

**Figure 5** The scaling relations of the hierarchy with cascade structure and related ubiquitous general

empirical observations

[**Note**: The 1/f noise is the special case of $1/f^{\beta}$ noise, and $\beta$ is the spectral exponent.]

The rank-size rule is the special case of the generalized Zipf's law, which can be expressed as

$$P_k = P_1 k^{-q}, \qquad (17)$$

where $q$ refers to the Zipf exponent. An interesting discovery is as follows. If $q \neq 1$, Zipf's law will be equivalent to and only to the generalized $2^n$ rule with $r_f = 2$ and $r_p = 2^q$ (This will be demonstrated in a companion paper). If $q=1$, Zipf's law can be reduced to the rank-size rule, which thus can be decomposed as two exponential laws with $r_f = r_p = N$ ($N$=2, 3, 4,…). Furthermore, an allometric scaling relation can be derived from equation (14). For example, city size can be measured not only by population but also by urban area. Substituting average area of cities in the $m$th class $A_m$ for the urban population size $P_m$ in equation (14) suggests the law of allometric scaling such as

$$A_m = a P_m^b, \qquad (18)$$

where $a$ refers to the proportionality coefficient, and $b$ to the scaling exponent (Chen, 2010).

The scaling decomposition and reconstruction of the rank-size rule can be generalized to the $1/f$ noise, which is defined by the following frequency-spectrum relation

$$S_f \propto \frac{1}{f}, \tag{19}$$

where $f$ denotes frequency, and $S_f$ is the spectral density, the symbol "$\propto$" means "is proportional to". The frequency can be written as $f=k/T$, where $k=1, 2, 3,…, T$ refers to the length of sample path. This suggests that equation (19) shares the identical form with equation (1). Thus equation (19) can be decomposed as two exponential laws for the hierarchy of spectral densities. The number law is the same as equation (2), while the density law is

$$E_m = N^{1-m} = E_1 r_e^{1-m}, \tag{20}$$

where $E_m$ is the average spectral density in the $m$th class, $E_1$ is the proportionality constant, and $r_e=E_m/E_{m+1}$ denotes the average density ratio. Based on equations (2) and (20), the scaling law can be reconstructed in the form

$$f_m = \eta E_m^{-\sigma}, \tag{21}$$

where $\eta=f_1 E_1$ is the proportional coefficient, and $\sigma=\ln r_f/\ln r_e=1$ is the scaling exponent.

A hierarchy is always associated with a network in space (Batty and Longley, 1994). Let's take cities as example to illustrate it. Comparing the experiment results listed in Table 1 and displayed in Figure 2 shows that when $N=3$, the subdivision effect of the rank-size distribution is the best for the limited classes. The hierarchy of $N=3$ corresponds to the central place network with $K=3$ (Christaller, 1966). A central place system is a hierarchy as well as a network following the scaling laws given above. A central place network can be characterized by the number law, equation (2), the size law, equation (14), and the hierarchical scaling law, equation (15) (Chen and Zhou, 2006).

Maybe the $N^n$ rule provides us with a new angle of view to understand the notion of "rank clocks" (Batty, 2006). At the micro-level, the rank clocks show cities rising and falling in size at many times and on many scales. However, at the macro-level, the hierarchical structure of national city-size distributions is always very stable and change little. If we regard different size classes as different "energy levels" of cities, change of status of an individual city is similar to transition of the city's energy level. However, when the Zipf exponent $q=1$, the $m$th level often has no essential difference from the $(m+1)$th level. In particular, no matter how individual cities change their positions, the total "energy" in each class is conservative and tries to keep constant. This suggests that hierarchical structure is independent of components, and a system is more important than its elements. On the other hand, no matter what number ratio ($N$=2, 3, 4, …) is taken, the number one never leave the top class, and the numbers two and three never leave the second class. Therefore, the first one or more cities always make an exception, and the top class often represents an extraordinary level.

# 5 Conclusions

This paper is involved with the method of applied mathematics, ideas from physics, and empirical evidences of geography. I am not sure to which classification it should belong, but I know it is important because this work shows a new approach to exploring varied scaling laws, which appear in many fields and have caused a broad research interest in the scientific circle. Especially, I proved the *geometric subdivision theorem* of harmonic sequence. The significance of this theorem is as follows.

Firstly, the rank-size rule is proved to be equivalent in theory to the $N^n$ principle. This suggests that the rank-size pattern might be a signature of hierarchical structure. The rank-size distribution with scaling exponent equal to 1 can be transformed into a self-similar hierarchy with fractal dimension $D$=1 according to the $2^n$ rule, $3^n$ rule, $4^n$ rule, and so on. Generally, we have the $N^n$ rule for the rank-size distributions. In this case, the scaling relation between rank and size can be decomposed as a pair of exponential laws and then reconstructed as the hierarchical scaling relation between size and number. Thus we can study a power law through a pair of exponential laws, which are simpler than a power law. What is more, we can get information of the rank-size distribution which can not be directly obtained through Zipf's law.

Secondly, the rank-size pattern suggests a special symmetrical structure. The rank-size rule is the special case of Zipf's law with scaling exponent varying from 0 to infinity. The rank-size rule is equivalent to the $N^n$ rule only for the rank-size distribution with negative exponent 1. If the scaling exponent gets departure from 1, the size distribution can only be converted into the hierarchy following the $2^n$ principle. This is revealing for us to understand the special scaling exponent $q$=1. The $N^n$ rule indicates that the hierarchy has no characteristic common ratio, and the scaling exponent 1 implies an arbitrary positive integer as common ratios for the hierarchy. Thus, different size levels can share the equal "political status" in a system except the first three ones.

Thirdly, the self-similar hierarchy is a magic framework of nature with many singular properties. The first is that it can act as a link between exponential laws and power laws, the second, many basic rules can be derived from this structure, and the third, it can be associated with a number of ubiquitous empirical observations. Especially, both the $1/k$ distribution and $1/f$ noise suggest the $N^n$ principle in the self-organized systems. A majority of the conclusion drawn from the rank-size rule can be generalized to the $1/f$ noise. In short, the hierarchical framework provides us with a revealing angle of view to understand how nature works. Moreover, the "magic" property of the harmonic sequence remains and is worthy to be mathematically studied in future.

## Acknowledgements:

This research was sponsored by the National Natural Science Foundation of China (Grant No. 40771061). The support is gratefully acknowledged. I would like to thank my mathematician friend, Juwang Hu, for assistance in mathematical demonstration.